\def\puncspace{\ifmmode\,\else{\ifcat.\C{\if.\C\else\if,\C\else\if?\C\else%
\if:\C\else\if;\C\else\if-\C\else\if)\C\else\if/\C\else\if]\C\else\if'\C%
\else\space\fi\fi\fi\fi\fi\fi\fi\fi\fi\fi}%
\else\if\empty\C\else\if\space\C\else\space\fi\fi\fi}\fi}
\def\SP{\let\\=\empty\futurelet\C\puncspace}

\def\h1{$h^{-1}$\SP}

\def\etal{{\it et al.\/}\ }

\def\eg{{\it e.g.\/}\rm,\ }
\def\lsim{~\rlap{$<$}{\lower 1.0ex\hbox{$\sim$}}}
\def\gsim{~\rlap{$>$}{\lower 1.0ex\hbox{$\sim$}}}
\def\void#1{{}}
\documentclass{aa}

\usepackage{graphicx}
\begin{document}

   \thesaurus{06     % A&A Section 6: Form. struct. and evolut. of stars
              (03.11.1;  % Cosmogony,
               16.06.1;  % Planets and satellites: general,
               19.06.1;  % Solar system: general,
               19.37.1;  % Stars: formation of,
               19.53.1;  % Stars: oscillations of,
               19.63.1)} % Stars: structure of.
   \title{ESO Imaging Survey}

   \subtitle{III. Multicolor Data near the South Galactic Pole }

\author {I. Prandoni\inst{1,2} \and  R. Wichmann\inst{1,3} \and L. da
Costa\inst{1} \and C. Benoist\inst{1,4} \and R. M\'endez\inst{1,5}
\and M. Nonino\inst{1,6} \and L.F. Olsen\inst{1,7} \and A. Wicenec\inst{1} \and S. Zaggia\inst{1,8} \and E. Bertin\inst{1,9,10}
\and E. Deul\inst{1,9} \and T. Erben\inst{1,11}  \and M.D.
Guarnieri\inst{1,12} \and I. Hook\inst{1} \and R. Hook\inst{13} \and
M. Scodeggio\inst{1} \and R. Slijkhuis\inst{1,9}
}

\institute{
European Southern Observatory, Karl-Schwarzschild-Str. 2, D--85748
Garching b. M\"unchen, Germany \and Istituto di Radioastronomia del
CNR, Via Gobetti 101, 40129 Bologna, Italy \and IUCAA, Post Bag 4,
Ganeshkhind, Pune 411007, India \and DAEC, Observatoire de
Paris-Meudon, 5 Pl. J. Janssen, 92195 Meudon Cedex, France \and Cerro
Tololo Inter-American Observatory, Casilla 603, La Serena, Chile
\and Osservatorio Astronomico di Trieste, Via G.B. Tiepolo 11, I-31144
Trieste, Italy \and
Astronomisk Observatorium, Juliane Maries Vej 30, DK-2100 Copenhagen, 
Denmark  \and
Osservatorio Astronomico di Capodimonte, via Moiariello 15, I-80131 Napoli,
Italy \and
Leiden Observatory, P.O. Box 9513, 2300 RA Leiden, The Netherlands \and
Institut d'Astrophysique de Paris, 98bis Bd Arago, 75014 Paris, France \and
Max-Planck Institut f\"ur Astrophysik, Postfach 1523 D-85748,  Garching b. 
M\"unchen, Germany \and
Osservatorio Astronomico di Pino Torinese, Strada Osservatorio 20, I-10025 
Torino, Italy \and
Space Telescope -- European Coordinating Facility, Karl-Schwarzschild-Str. 2, 
D--85748 Garching b. M\"unchen, Germany 
}

%   \author{EIS Team
%          \inst{1}
%          \and
%         C. Ptolemy\inst{2}\fnmsep\thanks{Just to show the usage
%          of the elements in the author field}
%          }

   \offprints{L. da Costa}

%   \institute{European Southern Observatorry, Karl-Schwazschild-Str. 2
%   Garching bei M\'unchen, Germany\\
%              email: 
%         \and
%             University of Alexandria, Department of Geography\\
%             email: c.ptolemy@hipparch.uheaven.space
%             \thanks{The university of heaven temporarily does not
%                     accept e-mails}
%             }

%   \date{Received September 15, 1996; accepted March 16, 1997}

   \maketitle
    
   \today

   \begin{abstract} This paper presents multicolor data obtained for a
1.7 square degree region near the South Galactic Pole (patch~B) as
part of the ESO Imaging Survey (EIS). So far the observations have
been conducted in $B, V$ and $I$, but are expected to be complemented
by observations in the U-band later in 1998. Object catalogs extracted
from single exposure images are 80\% complete down to $B \sim 24$, $V
\sim 23.5$ and $I \sim 22.5$, and once co-added should reach about 0.5
mag deeper. The data are being made public in the form of catalogs,
pixel maps, target lists and image ``postage stamps'', which can be
retrieved from the Web.  Counts of stars and galaxies and the angular
two-point correlation function of galaxies are computed and compared
to other available data to evaluate the depth and uniformity of the
extracted object catalogs. In addition, color distributions of stellar
objects are presented and compared to model predictions to examine the
reliability of the colors. The results suggest that the overall
quality of the catalogs extracted from the images is good and suitable
for the science goals of the survey.

\void{The results show that the data for patch B should be
useful for the selection of high-z candidate QSOs, to constrain
galactic models and to identify different stellar populations.}

   \end{abstract}

%
%________________________________________________________________

\section{Introduction}

The present paper is part of a  series presenting the data
accumulated by the public ESO Imaging Survey (EIS) being carried out
in preparation for the first year of regular operation of the VLT. As
described in previous papers (Renzini \& da Costa 1997, Nonino \etal
1998, hereafter paper I) the main science goal of EIS is the search for
rare objects such as clusters of galaxies, spanning a broad redshift
range, quasars at intermediate and high redshifts, high-redshift
galaxies and stars with special characteristics (\eg white dwarfs,
very low mass stars, brown dwarfs).  These goals have guided the
adopted survey strategy which, for EIS-wide, envisioned observations
in $V$ and $I$ to search for clusters and in four passbands ($U, B, V,
I$) over $\sim 1.7$ square degrees in a region near the South Galactic
Pole (EIS-wide patch B). The observations in $B, V$ and $I$ have
already been completed, while observations in the $U$-band are
expected to be carried out in the fall of 1998.

One of the main motivations for the multicolor survey has been the
identification of a large number of close line-of-sight, intermediate
redshift QSOs to study the three-dimensional distribution of
absorbers, using medium and high-resolution spectrographs (\eg UVES)
at the VLT. For this reason, a region near the SGP, where several QSOs
are known from previous studies, has been selected. However, the data
are also useful for galactic studies and for the identification of
rare stellar populations, with the survey having a unique combination
of depth and area coverage. The additional $V$-band images, which altogether
overlap the $I$ images over $\sim$ 70\% of the surveyed area in
patches A and B (15\% of the whole EIS-wide), are also useful in the
search of galaxy clusters (\eg Olsen \etal 1998, paper~V). Finally, it
is important to emphasize that the present data offer an excellent
opportunity to assess the complexity of efficiently handling large
volumes of multicolor data and extracting useful target lists. This is
a key element for the exploration of the full range of science offered
by the multicolor surveys envisioned for the new wide-field camera
(WFI@2.2m) at the ESO/MPIA 2.2m telescope at La Silla.

In section 2, the observations are described and the characteristics
of the multicolor data are presented. This section also presents the
extracted object catalogs and discusses their completeness and
reliability. In section~3 the catalogs are evaluated by comparison
with models and other data. Concluding remarks are presented in
section~4.

\section{Observations and Data Reduction}
\label{data}

\subsection {Observations}
\label{obs}

The observations of patch B were carried out over several months in
the period July 1997 to December 1997, using the red channel of the
EMMI camera on the 3.5m New Technology Telescope (NTT) at La Silla.
The red channel of EMMI is equipped with a Tektronix 2046 $\times$
2046 chip with a pixel size of 0.266 arcsec and a useful field-of-view
of about $9' \times 8.5'$. EIS uses a special set of BVI filters and
the response function of the system can be found in paper~I.

\begin{figure}[ht]
%\resizebox{\hsize}{!}{\includegraphics{seeinghistall.ps}}
\resizebox{\hsize}{!}{\includegraphics{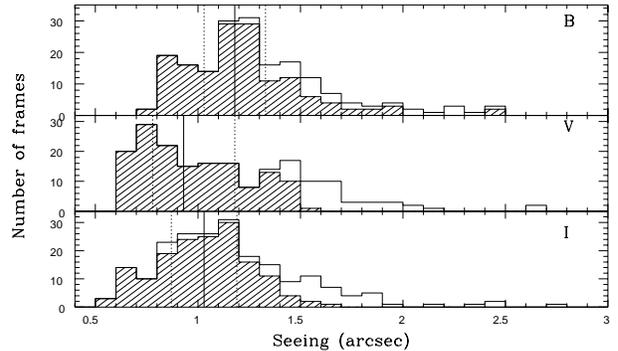}}               
\caption{Histogram of the seeing distribution for patch B obtained from 
all observed frames and only from the frames actually accepted for the
survey (shaded area). Vertical lines refer to 25, 50 and 75
percentiles of the accepted frames distribution. The three panels
refer to the three observed bands ($B, V, I$), as indicated in each
panel.}
\label{fig:seeinghistall}
\end{figure}

\begin{figure}%
%\resizebox{\hsize}{!}{\includegraphics{limmagdist.ps}}               
\resizebox{\hsize}{!}{\includegraphics{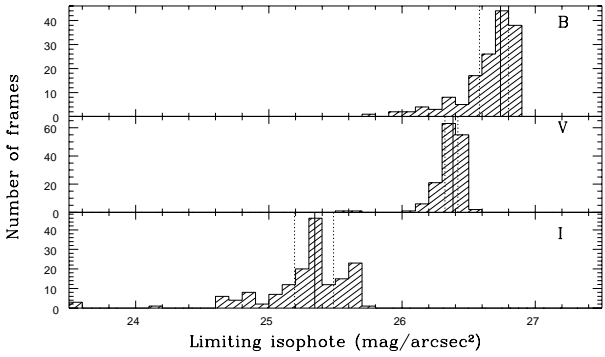}}               
\caption{Limiting isophote distributions from patch $B$ frames
actually accepted for the survey.  Vertical lines refer to 25, 50 and
 75 percentiles of the distributions. The three panels refer to the
 three observed bands ($B, V, I$).} \label{fig:mlim} \end{figure}

As described in paper I, the observations were carried out by a
sequence of overlapping exposures (hereafter refer to as even/odd
frames) 150 sec each, with each position on the sky being sampled at
least twice.  A total of 701 frames were obtained in the area with 200
in $B$, 282 in V and 219 in I bands.  Only 150 frames in each band
were required to cover the field but poor weather conditions required
several frames to be re-observed.  The strong variations in the
observing conditions can be seen in Figure~\ref{fig:seeinghistall}
which shows, for each band, the seeing distribution of all observed
frames. For comparison the shaded histograms show the seeing
distribution of the frames finally accepted, with the solid vertical
line in each panel indicating the median seeing for each band. The
$B$-band is the worst overall with a median seeing of 1.2 arcsec and
with a few frames extending to very large seeing ($\sim 2.5$
arcsec). In the analysis below 9 frames with seeing $\gsim 1.8$ arcsec
were discarded because of their incompleteness at faint magnitudes.
Figure~\ref{fig:mlim} shows, again for each band, the $1\sigma$
limiting isophote within 1 arcsec. The transparency of the nights also
showed significant variations especially for the $I$-band images, with
one frame reaching 21~mag/arcsec$^2$ (not shown in the figure). For
this frame, which was removed from analysis, the depth reached is
considerably shallower than the remaining frames leading not only to a
bright limiting magnitude but also to the detection of a significant
number of spurious objects. Other frames with bright limiting
isophotes have no significant impact in the analysis presented in
section~\ref{analysis}.

\begin{figure}[ht]
\resizebox{\hsize}{!}{\includegraphics{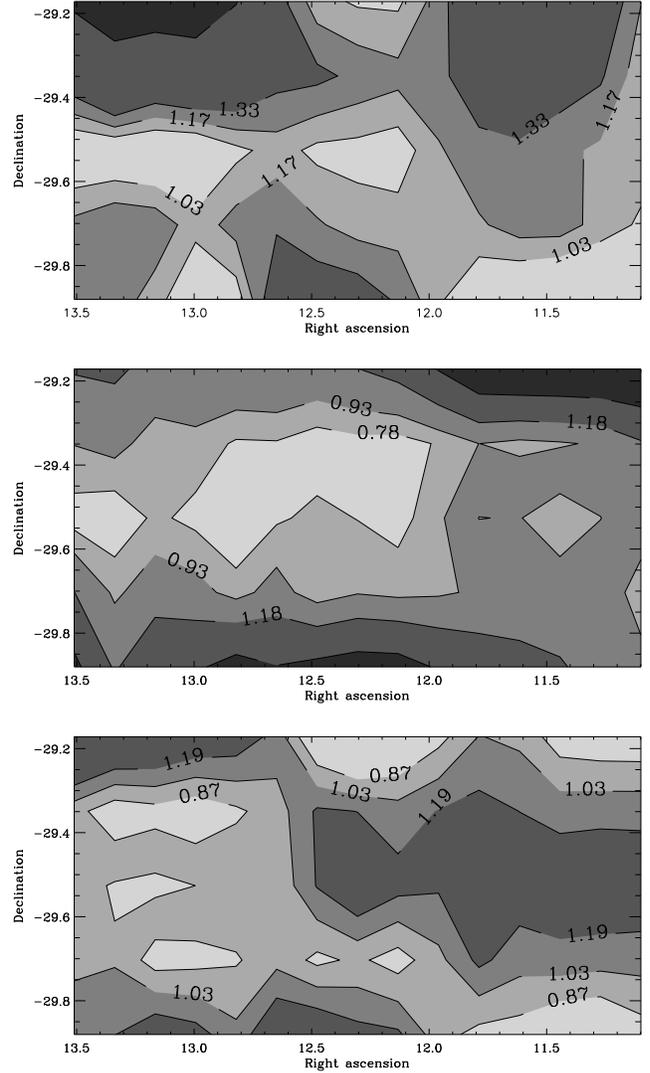}}               
\caption{Two-dimensional distribution of the seeing as measured
for patch B for all the accepted even frames. Contours
refer to 25, 50 and 75 percentiles of the distribution. The three panels 
refer to the three observed bands (B, V, I) from top to bottom. }
\label{fig:2dseeing}
\end{figure}

\begin{figure}[ht]
\resizebox{\hsize}{!}{\includegraphics{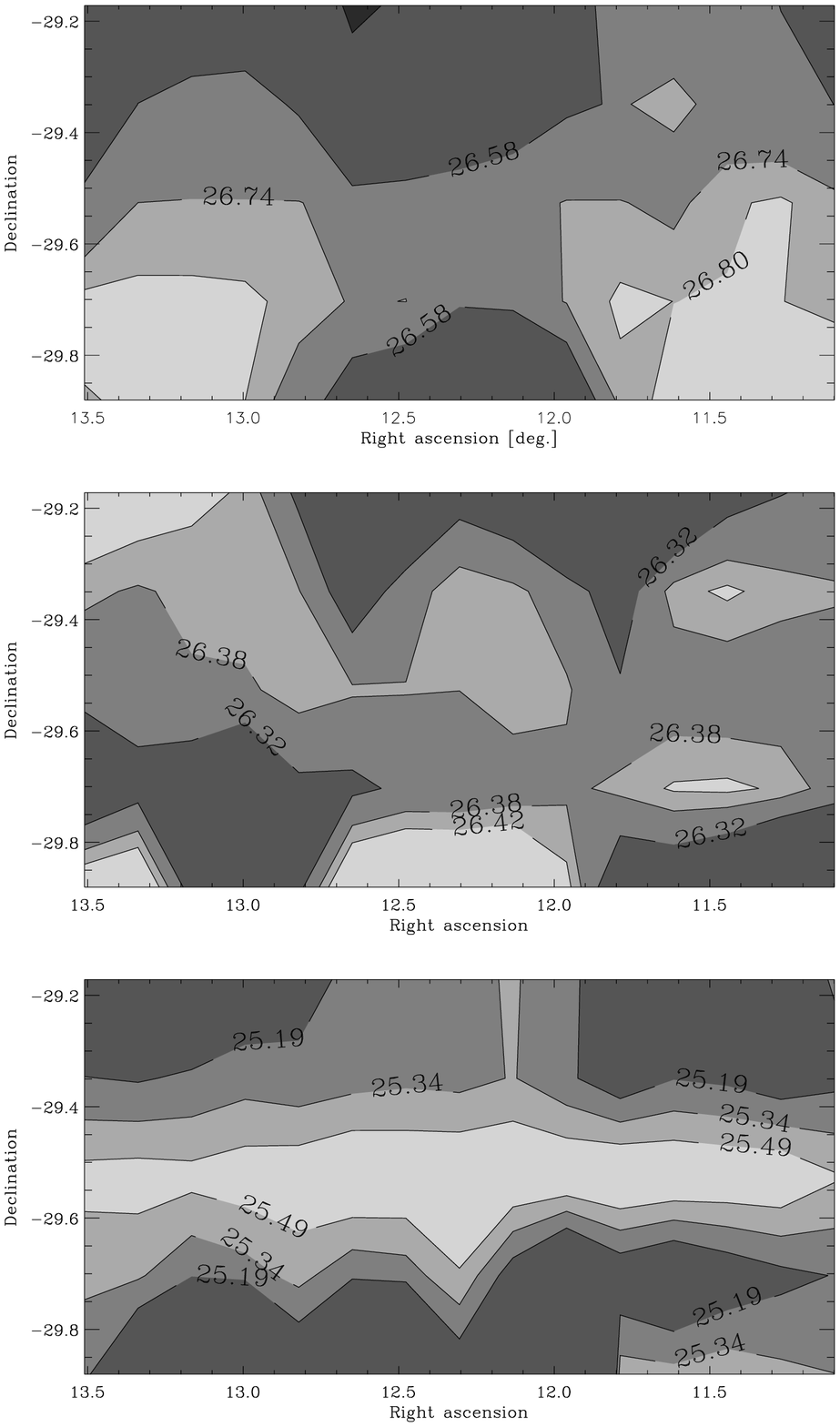}}               
\caption{Two-dimensional distribution of the 
limiting isophote as defined in the text estimated from the accepted
even frames for patch B.  Contours refer to 25, 50 and 75 percentiles
of the distribution. The three panels refer to the three observed
bands ($B, V, I$), from top to bottom.}
\label{fig:2dmlim}
\end{figure}

Figures~\ref{fig:2dseeing} and~\ref{fig:2dmlim} show, for each band,
the two-dimensional distribution of the seeing and limiting isophote
as determined from the even frames. Similar results are obtained for
the odd frames which alternate with the even ones. Such maps allow the
potential user of the derived catalogs to evaluate their reliability.
The final data is reasonably homogeneous with the median seeing in all
bands $< 1.2$ arcsec. However, some poor images do exist and for some
applications must be removed, as discussed above. The area that is
affected is $\lsim 0.1$ square degree.

Finally, it is worth mentioning that since the completion of paper~I,
the $V$ images for patch~A over an area $\sim 1.1$ square degrees have
also been reduced and are being made available together with the
catalogs extracted from them which are used below.

\subsection{Data Reduction}
\label{red}

The data were processed by the EIS pipeline being developed to handle
large imaging programs and described in detail in paper I. The
software development is still in progress with new functionalities
being constantly added to the pipeline as well as enhanced features in
Skycat driven by the survey needs, in particular to facilitate the
visual inspection of the target lists being produced. In addition, new
tools are being developed to handle color information, which adds a
new level of complexity especially in the preparation of object
catalogs. Computation of colors require reliable association of
objects detected in different passbands, which may be affected by the
seeing, astrometric errors, the morphology of the objects and the
performance of the de-blending algorithm.  One also needs a proper
definition for the measurement of colors for faint sources and upper
limits for non-detections. Currently, preliminary color catalogs are
being produced only for point-sources, which are being systematically
inspected to verify the catalogs and identify any peculiarities
(Zaggia \etal 1998). This is an important first step towards the
preparation of the final color catalog for patch B.  For instance,
during the visual inspection of objects selected by their peculiar
colors, it became evident that ghost images, observed near relatively
bright stars in the $B$ and $V$ images, contaminate the catalogs.

\subsection{Color Transformation}
\label{color}

\begin{figure}[ht]
\resizebox{\hsize}{!}{\includegraphics{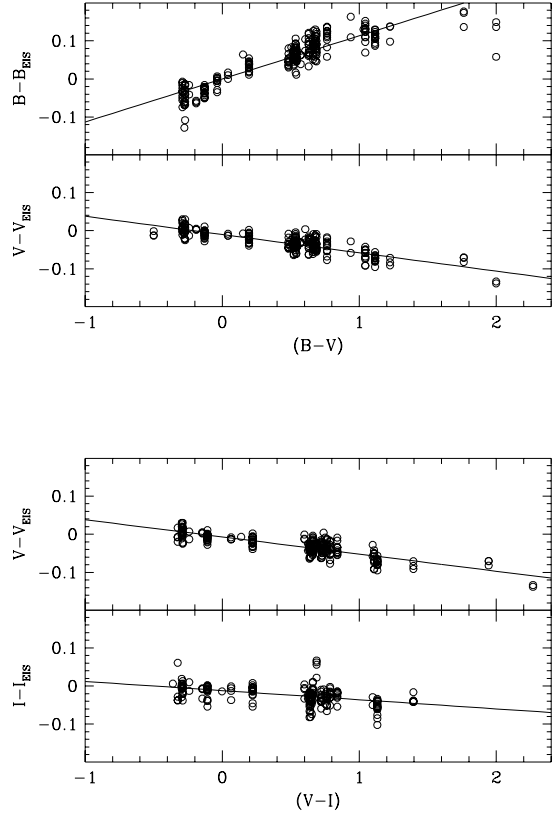}}               
\caption{Relation between the EIS and Johnson-Cousins system as a
function of color.  Shown are all the standard stars observed under
photometric conditions in the period July 1997-March 1998. }
\label{fig:color}
\end{figure}

Using all the standard stars observations carried out in photometric
nights at the NTT with the EIS filters, in the period July 1997-March
1998, the color transformation between the EIS and the Johnson-Cousins
systems has been determined. In Figure~\ref{fig:color} the observed
transformations for all the three bands are shown, as a function of
color in the Johnson-Cousins system. The fits are given by the
relations:

\begin{eqnarray}
B_{EIS} & = & B - 0.132 (B-V) \\
V_{EIS} & = & V + 0.047 (B-V) \\
V_{EIS} & = & V + 0.045 (V-I) \\
I_{EIS} & = & I + 0.036 (V-I) 
\label{conv}
\end{eqnarray}
Note that the transformation given here for $I_{EIS}$ is slightly
different from that determined in paper I. This is because more
standards have been included since and a more careful pruning of the
data has been performed.  The determination of color corrections
include 284 measurements in B, 255 in V and 209 in I, with the formal
errors in the color terms estimated to be $\lsim$ 0.02 mag in all
three bands. In general, the color term is small except for the
B-band. In this case the data also suggests a possible departure from
linearity at the red end. As a final note, it is worth mentioning that
in the process of examining all the standard stars observations,
errors in positions and the presence of variable stars in the Landolt
lists were found. A complete list of these problems will be reported
elsewhere.

\subsection{Calibration}
\label{calib}

The photometric calibration of the patch was carried out by first
bringing all frames to a common zero-point as determined from the
relative magnitudes of objects in overlap regions, within a
pre-selected magnitude range. This was done by a global least-square
fit to all the relative zero-points, constraining their sum to be
equal to zero. The internal accuracy of the derived photometric
solution is $\lsim 0.005$ mag (Paper~I). Second, absolute zero-points
are found for frames observed in photometric conditions.  The
zero-points for these frames were determined using a total of 36
frames of 7 fields containing standard stars taken from Landolt (1992
a,b), observed over 5 nights.  These frames were also reduced through
the pipeline, which identified the standard stars and measured
magnitudes through Landolt apertures automatically (see
paper~I). Altogether 148 independent measurements of standards were
used in the calibration.

Two solutions are then determined: one which computes a single
zero-point offset, based on the weighted average of the zero-points of
the calibrated frames, and the other using a first-order polynomial in
both right ascension and declination. Comparison with external data
suggests that a zero-point offset provides an adequate photometric
calibration for the entire patch.

\begin{figure}[ht] \resizebox{\hsize}{!}{\includegraphics{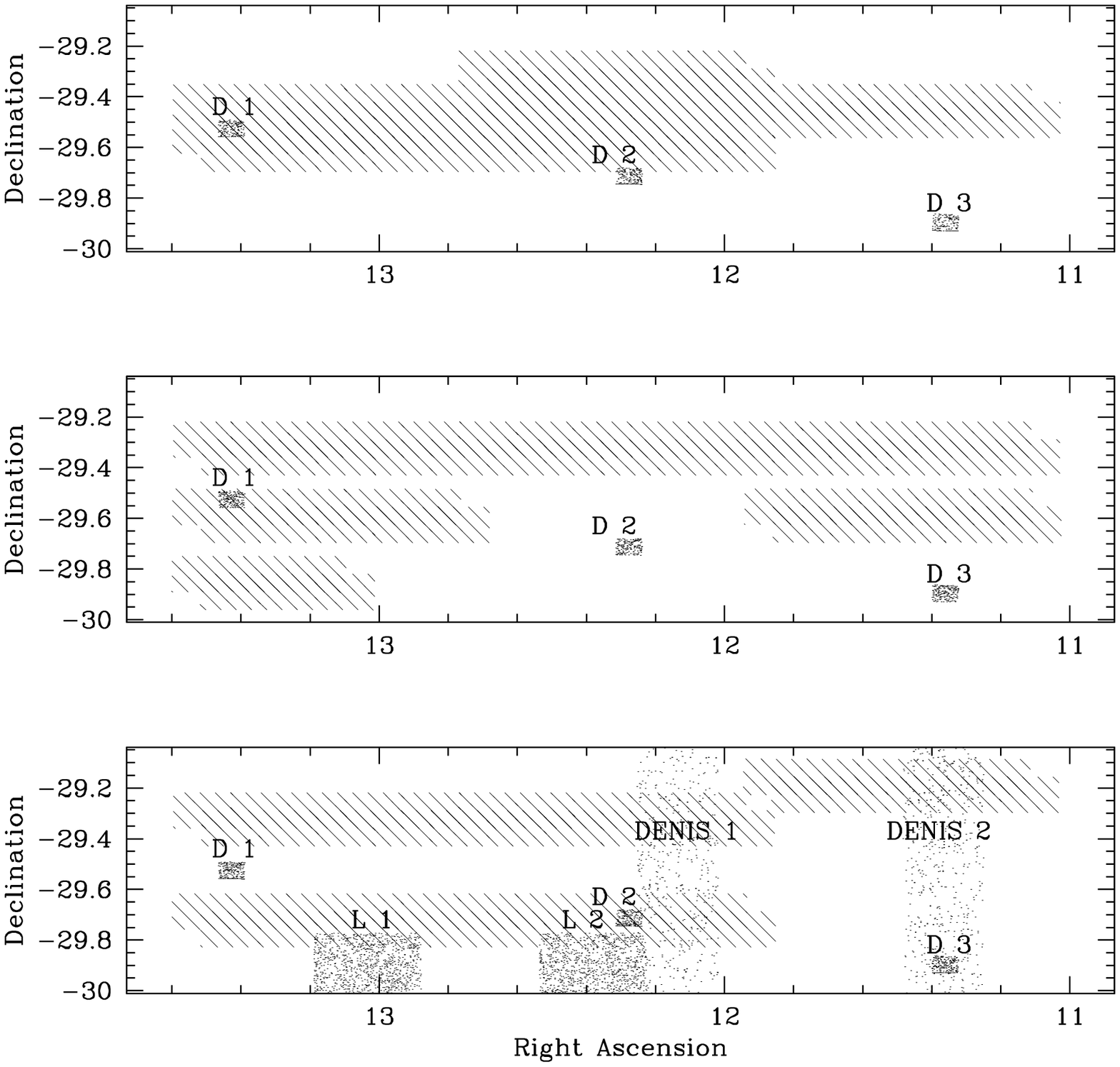}} 
              \caption{Distribution of frames obtained at the 0.9m
Dutch (D) telescopes at La Silla overlapping the surveyed region of
patch B. Also shown are parts of two DENIS strips that cross the field
and the Lidman \& Peterson fields (L) within the surveyed area. The
hatched area represents regions containing EIS frames observed under
photometric conditions.} \label{fig:overlaps} \end{figure}

External photometric data come from the Dutch 0.9m telescope at La
Silla and from overlaps with DENIS data and with frames taken by
Lidman \& Peterson (1996).  The regions of overlap of these data are shown in
Figure~\ref{fig:overlaps}. In the figure the regions observed under
photometric conditions are also indicated. Comparison of this figure
with its counterpart in paper I, demonstrates that the data for patch
B is clearly of superior quality with a much larger fraction of frames
taken under photometric conditions. Comparison with these external
data is important in order to look for possible gradients in the
photometric zero-point, introduced by the relative photometry which
implicitly assumes that there are no systematic errors in the
flatfield from frame to frame.

\subsection{Object Catalogs}
\label{cats}

During the processing of a patch through the pipeline, object catalogs
extracted from single frames are merged together into a ``patch''
catalog for each passband. This is the parent catalog which consists
of multiple entries of objects detected in overlapping frames.  For
each detection, the seeing and noise of the frame in which the object
was found is also stored.  The parent catalog is used to derive
different types of single-entry catalogs detected from 150~sec
exposures such as the odd/even catalogs described in
paper~I. Alternatively, it has also been used to derive a unique
catalog defined by examining the characteristics of the frames where a
given object was detected, saving only the entry associated with the
best seeing frame. Details regarding the methodology of association
will be described elsewhere (Deul \etal 1998).  From the flag
information available in the single-entry catalog, {\it filtered}
catalogs have been produced for analysis purposes.  The filtering is
required in order to eliminate truncated objects and objects with a
significant number of pixels affected by cosmics and/or other
artifacts. The parameters adopted in the filtering are the same as
those given in paper~I.

In general, this single-entry patch-wide catalog is the one used
below, while the odd/even are used to estimate the magnitude errors
directly from the data, by cross-identifying the objects. For
point-like sources, a preliminary attempt has also been made to
produce a color catalog combining the information of the catalogs
derived from each passband. Using the same association scheme
mentioned above, a cross-identification of objects is made and colors
are computed using the mag\_auto estimator of SExtractor (\eg Paper~I)
which should be adequate for point sources.  For non-detections in a
given band, 1$\sigma$ limiting magnitudes are computed from the seeing
and noise properties of the best seeing frame available at the
expected position of the object. Even though still rudimentary, this
derived catalog serves for verification purposes and for a first cut
analysis of the data.  The final color catalog will only be derived
from the co-added images. In this case, colors will be computed using
the detection area in each band as determined from the limiting
isophote defined either in one band (\eg $I$) or the summed images of
different bands (\eg $V+I$). Even though the required software is
available it is only now being integrated into the pipeline.

It is important to emphasize the complexity of handling and merging
information extracted from different passbands. For instance, each
object may have a different SExtractor stellarity-index which may
impact the galaxy/star classification, close pairs may be de-blended
in one passband and not in another, depending on the seeing.  Clearly
a complete description of all the possible pitfalls and the overall
performance of the software is beyond the scope of the present
paper. The current work also shows the shortcomings of handling
catalogs and points out the need for the implementation of an object
database with a flexible user interface to allow for the full
exploration of the data by different groups.

\begin{figure*}[t]
\resizebox{14cm}{!}{\includegraphics{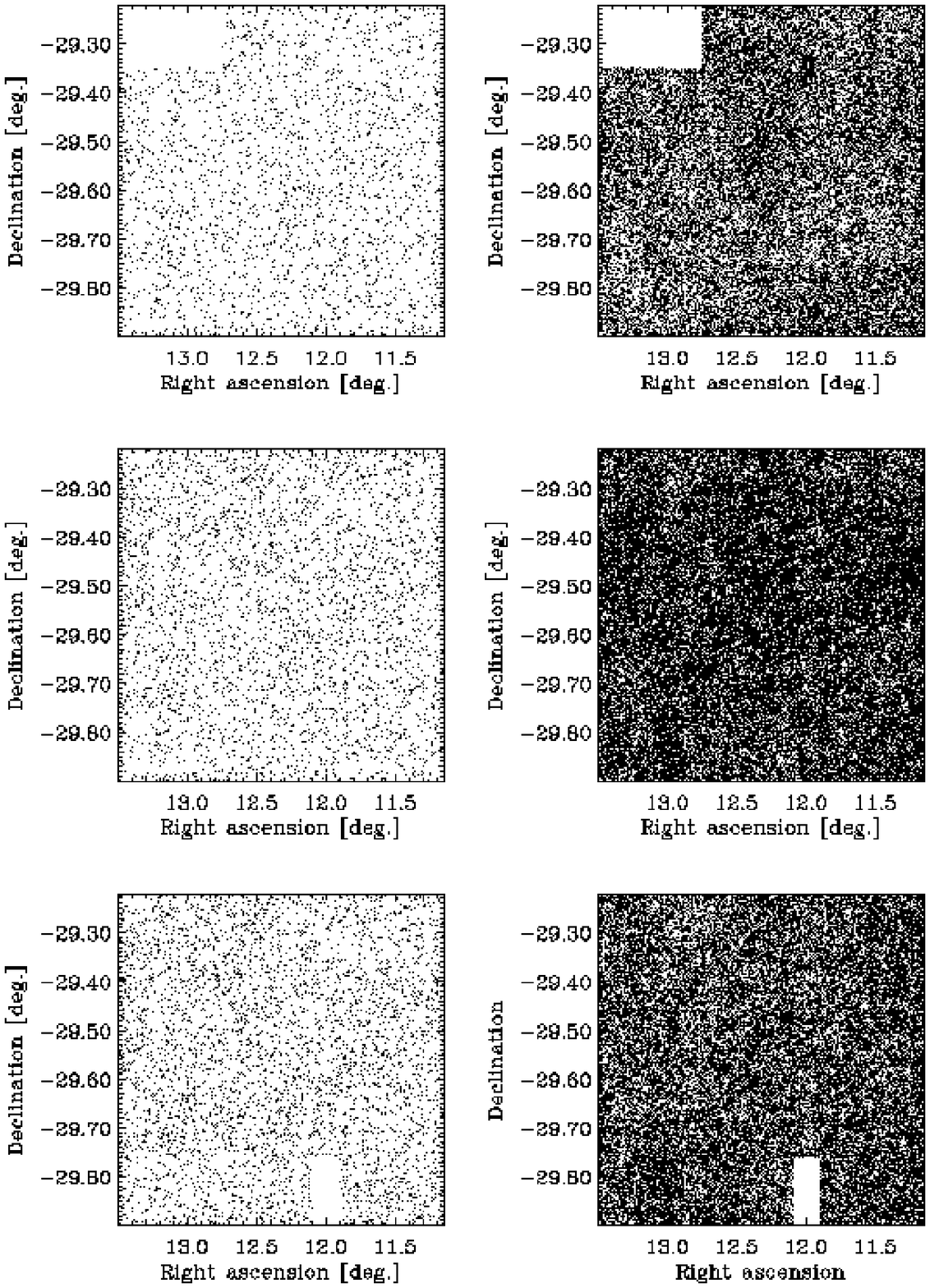}}               
\caption{Projected distribution of stars (left panel) and galaxies
(right panel) detected in the passbands $B$ (top panels),$ V$ (middle
panels) and $I$ (bottom panels). The limiting magnitude corresponds to
the star/galaxy classification limit for stars and to the estimated
80\% completeness limit for galaxies (see text). Frames taken
under extremely large seeing or with large extinction have been
eliminated.}
\label{fig:proj_dist}
\end{figure*}

For the purposes of the present paper galaxies are objects with
stellarity index $<0.75 $ if brighter than the star/galaxy
classification limit ($B$=22, $V$=22, $I$=21) or any object,
regardless of the stellarity index, fainter than this limit. Stars are
objects with stellarity index $\ge~0.75$.  Note that this definition
leads to some cross-contamination but it has no significant impact on
the conclusions. In evaluating the data in section~\ref{analysis} the
derived star and galaxy catalogs were, for simplicity, trimmed at the
edges and the bad frames discussed above were removed. After trimming
the areas covered are: 1.3, 1.4 and 1.37 square degrees in $B, V, I$,
respectively. The corresponding two-dimensional distributions of
stars, down to the star/galaxy classification limits, and galaxies,
down to estimated 80\% completeness limits, are shown in
figure~\ref{fig:proj_dist}.  The total number of objects in these
plots are: 2290 stars brighter than $B=22$ and 33133 galaxies brighter
than $B=24$; 3378 stars brighter than $V=22$ and 58590 galaxies
brighter than $V=24$; and 4297 stars and 44546 galaxies brighter than
$I=21$ and $I=22.5$, respectively.\void{These limits correspond roughly to
3$\sigma$ detections.} Recall that the distribution shown is for the
``best'' catalog. If one wishes to work with the odd/even catalogs
their distribution have to be examined in the same way. In order to
avoid extraneous colors due to bad data in one or more bands, the
color catalog examined in section~\ref{analysis} corresponds to the
common areas covered in the different passbands and has a total area
of 1.27 square degrees.

\subsection{Completeness, Contamination and Magnitude Errors}

\begin{figure}[ht] \resizebox{\hsize}{!}{\includegraphics{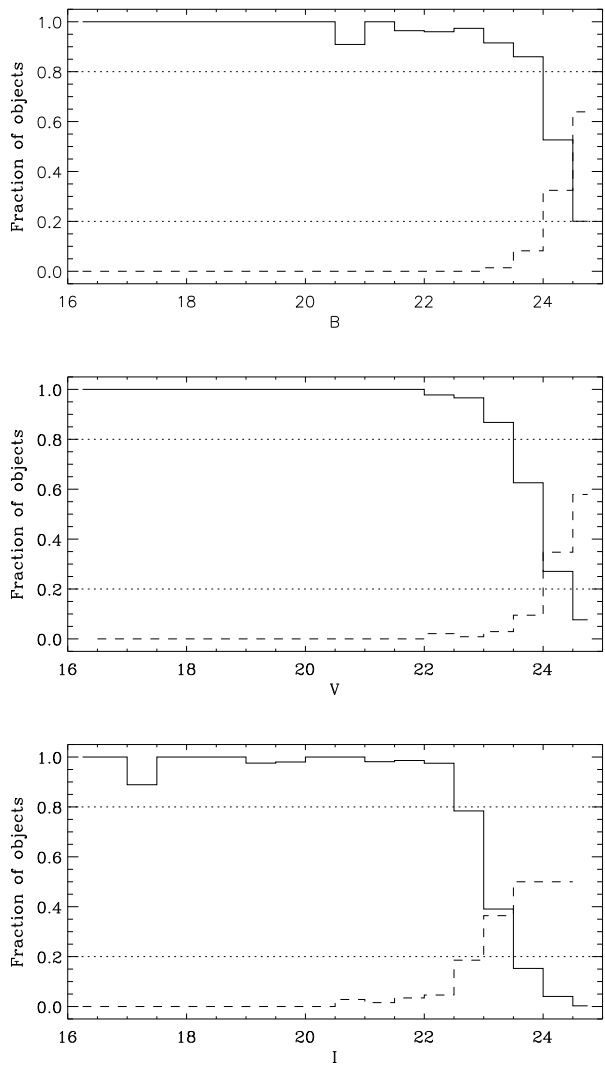}}
               \caption{Completeness (solid line) and expected
contamination by spurious objects (dashed line) in the EIS catalogs for
the different passbands considered. The computation of these quantities
are described in the text.} \label{fig:completeness} \end{figure}

As in paper~I, the completeness of the derived catalogs has been
established by using a single field where several exposures have been
made over the period of observations of patch B. The corresponding
catalogs in each passband were then compared with that of a typical
frame with an exposure time of 150~sec.  A total of 9 exposures in B
band, 6 exposures in V band and 8 in I band, were selected discarding
others taken in less favorable conditions. These exposures were
coadded and the derived catalogs are at least one magnitude deeper
than the typical survey frame. Comparing the identified objects in a
single exposure frame with those extracted from the co-added images
one can derive the expected completeness for typical 150 sec survey
frames. The results of these comparisons are shown in
figure~\ref{fig:completeness} showing that the catalogs are 80\%
complete at $B \sim 24$, $V \sim 23.5$ and $I
\sim 22.5$. These limits should apply for survey frames with a seeing
close to the median seeing. Also shown in the figure is the number of
false positives computed from the fraction of objects identified in
the single exposure frames which were not detected in the co-added
image.

%\begin{figure}[ht]
%\resizebox{\hsize}{!}{\includegraphics{reliability.ps}}               
%\caption{Completeness of the EIS catalogs for the different passbands
%considered. }
%\label{fig:reliability}
%\end{figure}

\begin{figure}[ht]
\resizebox{0.8\hsize}{!}{\includegraphics{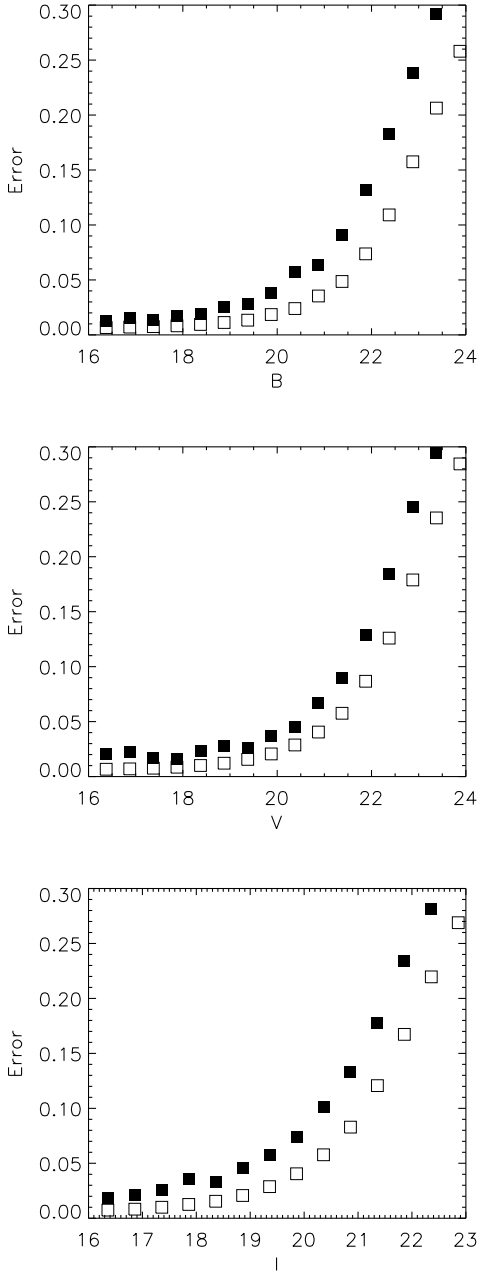}}               
\caption{Comparison  between the estimated error in the magnitudes from the
odd/even comparison (solid squares) and the SExtractor estimates (open
squares).}
\label{fig:errors}
\end{figure}

In order to estimate the accuracy of the magnitudes the odd and even
catalogs were compared and a lower limit estimate of the photometric
errors can be obtained from the repeatability of the magnitudes for
paired objects. The estimated errors from this comparison are given in
Figure~\ref{fig:errors}, which shows that in the interval $16<I<20.5$
they range from 0.02 to 0.1 mag, reaching 0.3 mag at $I \sim
23$. Similar values are found for $B$ and $V$ brighter than $\sim
24$.  These values correspond well to those estimated from
SExtractor.

\void{
compare with the
estimates given by SExtractor.
Figure~\ref{fig:errors} shows the magnitude difference
of these objects as a function of magnitude.  The standard deviation
of the magnitude differences in the interval $16 < I < 20.5$ ranges
between 0.02 and 0.1 mag, reaching 0.3 mag at $I \sim 23$.
Figure~\ref{fig:errors} also shows a comparison between the errors
determined from the magnitude difference shown above (divided by
$\sqrt{2}$) and the SExtractor error estimates based on photon
statistics. As can be seen, SExtractor provides reasonable error
estimates over the interval of interest. At bright magnitudes
photometric errors are dominated by effects such as flatfield errors,
image quality, intrinsic stability of the MAG\_AUTO estimator and
relative photometry.}

\begin{figure}[ht]
%\resizebox{\hsize}{!}{\includegraphics{eis_vs_lidman.tex.ps}}
%
\resizebox{\hsize}{!}{\includegraphics{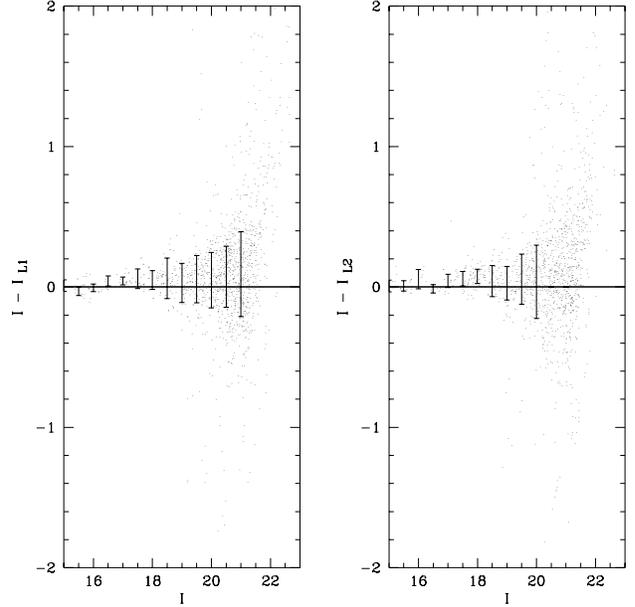}}               
\caption{Comparison of the EIS data in I-band with Lidman \& Peterson
(1996) catalog for the two fields in common. Also shown are the
mean and the rms in 0.5~mag bins.}
\label{fig:lidman}
\end{figure}

It is also of interest to obtain a completely independent estimate of
the errors in the magnitudes. This can be done by comparing the
objects detected by EIS with those found by Lidman \& Peterson (1996),
who have used a different object detection algorithm. This is shown in
figure~\ref{fig:lidman} where the objects detected in two separate
fields (see figure~\ref{fig:overlaps}), with 1354 and 1299 objects
each, are compared.  Note that because of differences in the
astrometry this comparison was done using the astrometric solution
found by the EIS pipeline but the magnitudes as determined in the
original catalog.  Even adopting this procedure misidentifications are
still present, leading to significant magnitude differences over the
entire magnitude range. Nevertheless, the zero-point offset is
typically $\sim$ 0.05~mag for $I < 21$, consistent with the zero-point
correction proposed by these authors to bring their measurements into
the Johnson-Cousins system. Beyond this limit, the Lidman \& Peterson
catalog becomes increasingly incomplete leading to a biased offset.
The scatter in this comparison is less than 0.3 mag down to $I \lsim
22$, consistent with our internal estimates if one attributes
comparable errors to the Lidman \& Peterson measurements. The
zero-point offset and the scatter of the magnitude differences is the
same for the two fields considered, suggesting that there are no
strong gradients in the photometric zero-point of the patch, at least
on a 0.5 degree scale, corresponding to the separation of the two
Lidman \& Peterson fields.

\begin{figure}[ht]
\resizebox{\hsize}{!}{\includegraphics{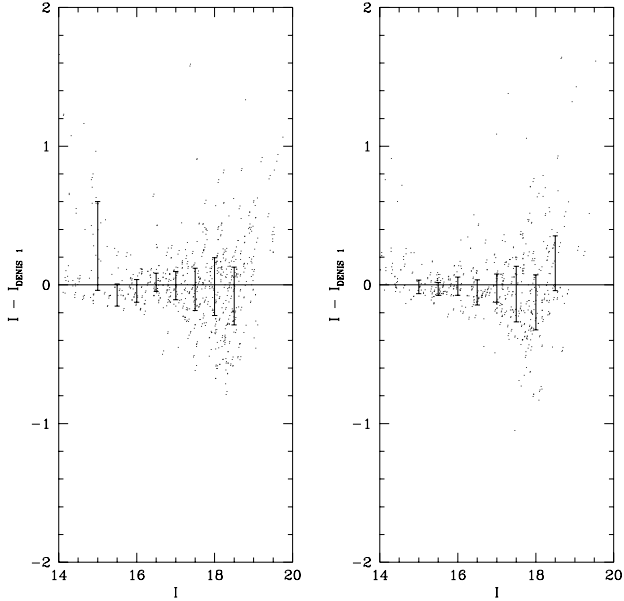}}               
\caption{Comparison of the EIS I-band magnitudes with those measured
by DENIS for the two strips that overlap patch B. Also shown are the
mean and the rms in 0.5~mag bins. }
\label{fig:denis}
\end{figure}

In order to further investigate possible systematic errors in the
photometric zero-point over the scale of the patch, the EIS catalogs
were also compared with object catalogs extracted from the two DENIS
strips that cross the survey region (see
figure~\ref{fig:overlaps}). This allows one to investigate the
variation of the zero-point as a function of right ascension and,
especially of declination. The results are shown in
figure~\ref{fig:denis}.  The domain in which the comparison can be
made is relatively small because of saturation of objects in EIS at
the bright end ($I\sim16$) and the shallow magnitude limit of DENIS
($I\sim18$). Still, within the two magnitudes where comparison is
possible one finds a roughly constant zero-point offset of less than
0.02~mag for both strips and a scatter that can be attributed to the
errors in the DENIS magnitudes (Deul 1998).

\begin{figure}[ht]
\resizebox{\hsize}{!}{\includegraphics{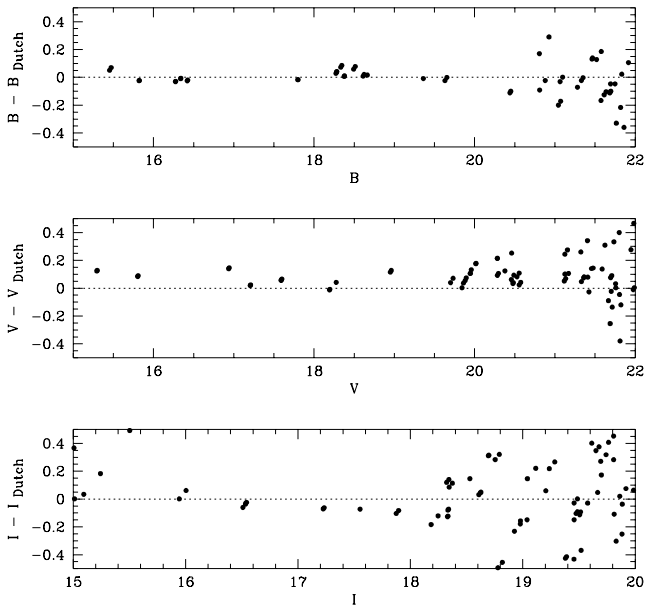}}               
\caption{Comparison of the EIS magnitudes in $B, V$ and $I$ (top to
bottom) with those measured from observations of the Dutch 0.9m
telescope. Objects in the three fields available have been combined.}
\label{fig:dutch}
\end{figure}

Finally, similar comparisons can be made between the EIS magnitudes
and those measured from the images obtained at the 0.9m Dutch
telescope at La Silla, in this case, for all three
passbands. Figure~\ref{fig:dutch} shows these comparisons, combining
all the three fields that overlap patch~B.  Even though the total
number of objects is relatively small (112 in $B$, 180 in $V$ and 204
in $I$) preventing an accurate comparison, one finds a reasonable
agreement in the zero-point and a scatter that can be accounted by
magnitudes errors in the Dutch data ($\sim 0.2$ at $B=21.5$, $V=21.5$
and $I=20$). The observed zero-point offset between the Dutch and EIS
data ($\sim 0.04$ in $B$, $\lsim$ 0.1 in $V$, $\lsim 0.02$ in $I$) can
be explained by the color term corrections required for the EIS and
Dutch measurements to bring both measurements into the Cousins
system. As the fields are well separated in right ascension, this
result gives further evidence that there are no significant gradients
in the photometric zero-point in any of the passbands.

In summary, comparison of the EIS magnitudes with available external
data shows no indication of gradients in the photometric zero-point of
the patch.

\begin{figure}[ht]
%\resizebox{\hsize}{!}{\includegraphics{starcounts.new.ps}}               
\resizebox{\hsize}{!}{\includegraphics{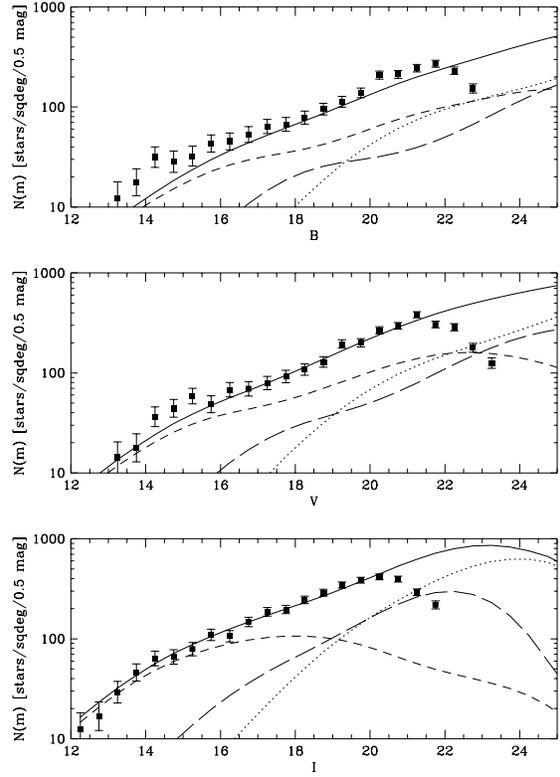}}               
\caption{The EIS differential star counts versus magnitude compared to the 
galactic model predictions (solid line), as described in the text. The
model includes an old disk population (dashed line), a thick disk
component (long-dashed line) and a halo (dotted line).  }
\label{fig:star_counts}
\end{figure}

\section{Data Evaluation}
\label{analysis}

Although this paper does not intend to interpret the data, some basic
statistics are computed to evaluate the overall performance of the EIS
pipeline in translating images into useful scientific products. For
this purpose, the stellar and galaxy samples extracted in each
passband and the preliminary color catalog for point-sources are
compared below with other available data and model predictions.

\subsection{Point-like Sources}

%\subsubsection{Comparison with Galactic Models}

Figure~\ref{fig:star_counts}, shows the comparison of the star counts
for patch B derived using the stellar sample extracted from the object
catalogs produced in each passband (section~\ref{cats}), with the
predicted counts based on a galactic model composed of an old-disk, a
thick disk and a halo. The star- and color-counts presented in this
section have been computed using the model described by M\'endez and
van Altena (1996), using the standard parameters described in their
Table~1. It is important to emphasize that no attempt has been made to
fit any of the model parameters to the observed counts. The model is
used solely as a guide to evaluate the data and, as can be seen from
figure~\ref{fig:star_counts} one finds a remarkable agreement with the
predicted counts down to the magnitude where the star-galaxy
separation is expected to become unreliable. The excess in counts seen
at the bright end is due to the saturation of brighter objects
($V\lsim16$).

\begin{figure*}[ht]
\resizebox{0.75\hsize}{!}{\includegraphics{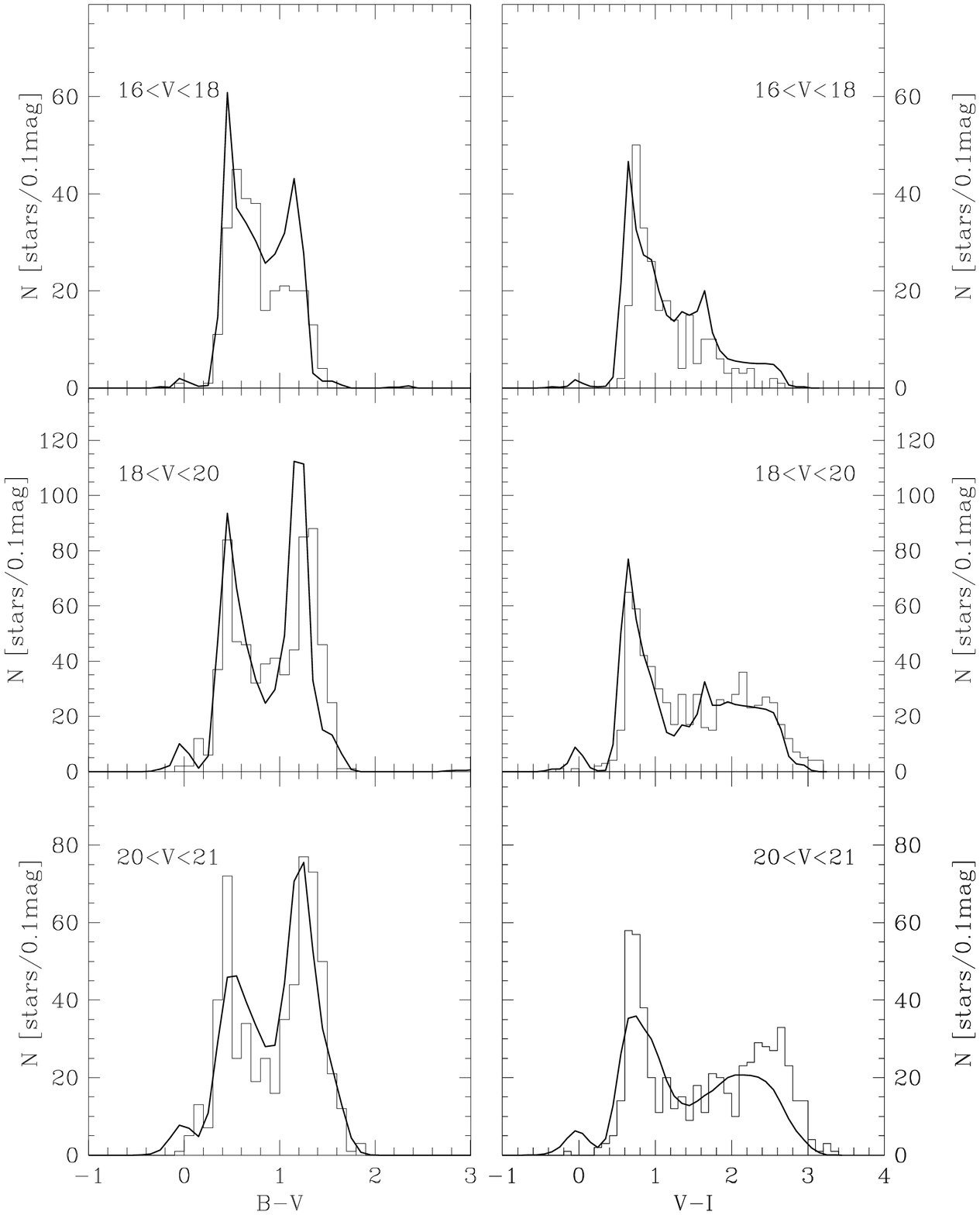}}               
%\resizebox{\hsize}{!}{\includegraphics{colhist_slices.ps}}               
\caption{
Color distribution for point-sources for different magnitude intervals
as indicated in each panel. Also shown are predictions for the
galactic model of M\'endez \& van Altena (1996).
}
\label{fig:colorbin}
\end{figure*}

Using the preliminary color catalog for point sources the observed
color distribution of stars brighter than $V=21$ is compared to model
predictions in figure~\ref{fig:colorbin} over three ranges of
magnitude as indicated in each panel. The model computes starcounts in
$B, V$ and $I$ by adopting a series of color-magnitude diagrams
appropriate for the disk, thick-disk and halo of our Galaxy. In order
to output predicted counts in the natural passbands of the EIS survey,
the transformation given by equations (1)-(4) have been used to
convert from the EIS magnitudes to the Johnson-Cousins system in such
a way that the predicted counts are actually evaluated in the EIS
passbands and are convolved using the error model given in
figure~\ref{fig:errors}. As can be seen from
figure~\ref{fig:color}, the number of red ($(B-V)>1.2$) standard
stars defining the color transformation is very small, so that one
might expect possible discrepancies between the observed and predicted
counts, particularly for redder colors. One way of overcoming this
would be to use synthetic star colors using the system response
functions given in paper~I. However, for the purposes of describing
the usefulness of the data, the current calibration is
sufficient. Considering that none of the model parameters have been
adjusted to fit the present data, the good agreement of the model to
the observed counts in both $(B-V)$ and $(V-I)$ is remarkable,
although some discrepancies can also be readily seen.

At the brightest magnitude bin one sees a deficit of red objects
relative to the model predictions in both ($B-V$) and $(V-I)$,
especially in the former, which is due to saturation effects. Note
that objects with saturated pixels have been discarded from the color
catalog.

In the range $18<V<20$, the color-counts are known to split into two
major peaks, each sampling a different stellar population (Bahcall
1986). The blue peak at $(B-V)\sim0.5$ is due to halo stars near the
turnoff ($M_V\sim+4$), located at few kpc from the Galactic plane. The
red peak at $(B-V)\sim1.3$ is due to faint M-dwarf stars from the
disk, located at less than 1~kpc from the Sun. Note the small relative
offset ($\sim0.1$~mag) between the observed and predicted location of
the red peak. As pointed out above this may be due to, and is
consistent with, the departure of the color term from the linear
correction adopted for objects with $(B-V)>1.2$. Note that the
agreement is much better in $(V-I)$ for which the contribution from
color terms are expected to be negligible.

Traditionally, the observed splitting in the color peaks has been used
to determine the local normalization of halo stars in the solar
neighborhood.  Note in this context the difference in the amplitude of
the counts in the blue peak in the magnitude range $20<V<21$, which is
seen in both $(B-V)$ and $(V-I)$. Most photometric surveys at faint
magnitudes (\eg Reid and Majewski 1993) have relied on pencil-beam
surveys covering a small fraction of a degree. Therefore, the number
of observed objects per bin has been very small, leading to large
uncertainties in the derived model parameters. The EIS sample,
covering $\sim 1.3$ square degrees, represents a significant
improvement and may allow for a better determination of these
parameters.

Finally, it is interesting to point out the existence of a population
of blue objects, in particular, the suggestion of a peak at
$(B-V)\sim0.15$ observed at faint magnitudes (20$<V<21$). This peak
does not match the location and the amplitude of the peak predicted by
the white dwarf population assumed in the model. Instead the observed
blue objects could consist of a mix of white dwarfs, blue horizontal
branch stars or perhaps halo field blue stragglers. Further
investigation on the nature of these objects seems worthwhile.

The results demonstrate that the stellar color catalog being produced
is by and large consistent with model predictions and the observed
differences may possibly point to deficiencies in the model which
should be further investigated by interested groups.  Although
primarily driven by other goals, the above discussion shows that the
EIS data is also useful for galatic studies.

\void{
Another potentially interesting result is the suggestion for presence
of the (expected) population of disk white dwarfs at $B-V \sim 0$,
even though the observed amplitude is significantly smaller than that
predicted by the model. In general, the space density of these objects
is so small ($\rho_{WD}
\sim 0.0034 \,$ stars/pc$^3$ for $M_V < +17$), that they only appear
in sufficient numbers on large-area surveys. The discrepancy between
the observed and predicted counts can perhaps be used to constrain the
scale-height of their distribution. Another interesting point is that
at $V> 18$ there appears to be a noticeable excess in the counts at
$(B-V) \sim 0.15$, which is {\it not} expected for WDs. An alternative
is that this excess could be associated with either blue horizontal
branch stars, which do not seem to be well represented in the model,
or perhaps to halo field blue stragglers.}

\void{A basic issue that could
be immediately be addressed with the EIS data is the scale-height of
these objects. Since the bright portion of the luminosity function for
WDs is well known (Liebert, Dahn and Monet 1988, although see the
recent claim to the opposite on Wood and Oswalt 1998), the only model
parameter that remains uncertain in these predictions is their
scale-height. The current model computations assume a value of 325~pc,
although values in the range 250 to 450 pc have been suggested (Boyle
1989, Villeneuve et al. 1995).}

\void{ In the range $16 \leq V < 21$
there are about 20 very likely WD candidates in the EIS catalog. It
would be extremely interesting to perform a spectroscopic follow-up of
these blue objects to determine their true nature, and to study issues
such as the ratio of DA to non-DA WDs as a function of color
($T_{eff}$) on a complete and well-defined sample. These studies bear
upon a wide range of stellar problems, from theories of WD formation
and cooling, to the age of the Galactic disc (Leggett, Ruiz and
Bergeron 1998).}

\void{
In general, analysis of the star counts use the color distributions in
different ranges of magnitudes. This is shown in
Figure~\ref{fig:colorbin}. In the figure we also show the predicted
model of Reid \& Majewski (1993) produced to fit the data for the
North Galactic Pole. There is a remarkable agreement between the data
and the model predictions, except at the faintest bin which is
probably due to incompleteness of the sample. Based on these results
we estimate that the data for patch B contains some $\sim$ ???? halo
stars.}

\subsection {Galaxies}

%\subsubsection {Galaxy Counts}

\begin{figure}
\resizebox{9cm}{!}{\includegraphics{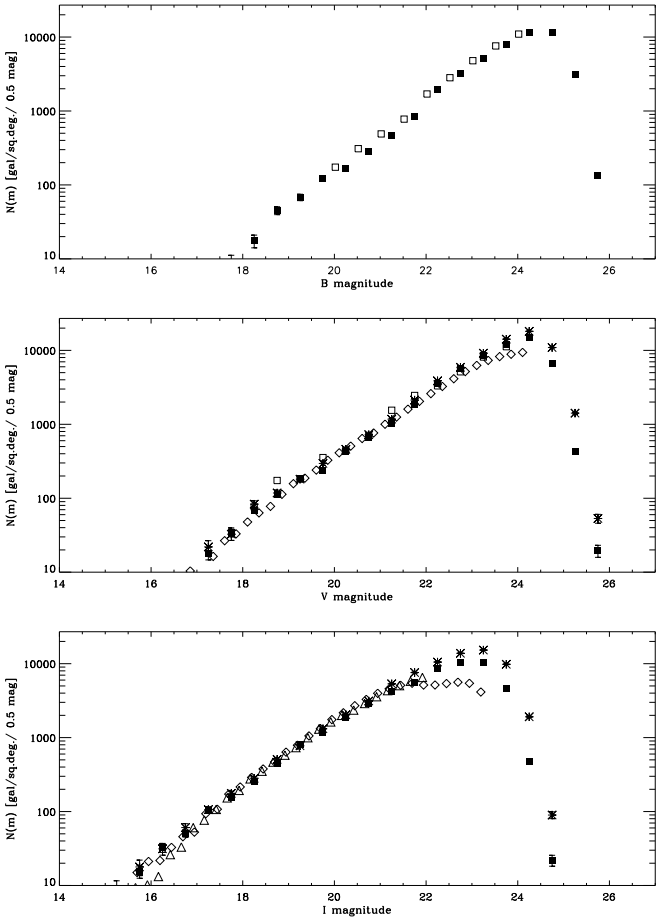}}
\caption{EIS patch B galaxy counts (filled squares) in different
passbands ($B,V,I$, from top to bottom) compared to the counts
obtained by: Lidman \& Peterson (triangles), Postman \etal (diamonds)
and Arnouts \etal (1997) (open squares), as provided by the
authors. Also shown are the counts obtained in patch A (stars) for the
$V$ and $I$-band.  The counts from other authors have been converted
to the Johnson-Cousins system, as described in the text.}
\label{ncounts_gal}
\end{figure}

In order to evaluate the quality and the depth of the galaxy samples,
galaxy counts in the different passbands are shown in
figure~\ref{ncounts_gal} and compared to those determined for patch~A
and by other authors as indicated in the figure caption.  In these
comparisons the $I$ magnitudes of Lidman \& Peterson (1996) have been
shifted by +0.04~mag and those measured by Postman \etal (1996) by
-0.43~mag to bring them into the Johnson-Cousins system. A small
correction (-0.02~mag) has also been applied to the $V$ counts of
Postman \etal. No corrections were made to the Arnouts \etal (1997)
data. As can be seen there is a remarkable agreement between the EIS
counts and those obtained by other authors. They are also consistent
with the counts determined from patch A, down to $V\sim24$ and
$I\sim22.5$. As emphasized in Paper~I even for single exposures EIS
reaches fainter magnitudes than previous data used for cluster
searches.

\void{
 the galaxy counts derived from the EIS catalogs are compared to those
of Lidman \& Peterson (1996) and Postman \etal (1996). The $1 \sigma$
error bars are computed as above.  There is a remarkable agreement
between the EIS galaxy-counts and those of the other authors. The
slope of the EIS counts is found to be $0.43 \pm 0.01$. Also note that
the EIS counts extend beyond those of Postman \etal (1996) even for
the counts derived from single frames. The galaxies have been defined
to be objects with a stellarity index $<$ 0.75 for $I < 21$ and all
objects fainter than $I = 21$. At this limit galaxies already
outnumber stars by a factor of $\sim$ 3.  }

%\subsubsection{Angular Correlation Function}

The overall uniformity of the EIS galaxy catalogs can be examined
using the two-point angular correlation function, $w(\theta)$.
Indeed, $w(\theta)$ is a very efficient tool for detecting any kind of
artificial patterns (such as a grid with scale comparable to an EMMI
frame) or possible gradients in the density over the field (which
could result from  large-scale gradients of the photometric
zero-point). Departures from uniformity should affect the correlation
function especially at faint magnitudes.

\begin{figure}
\resizebox{9cm}{!}{\includegraphics{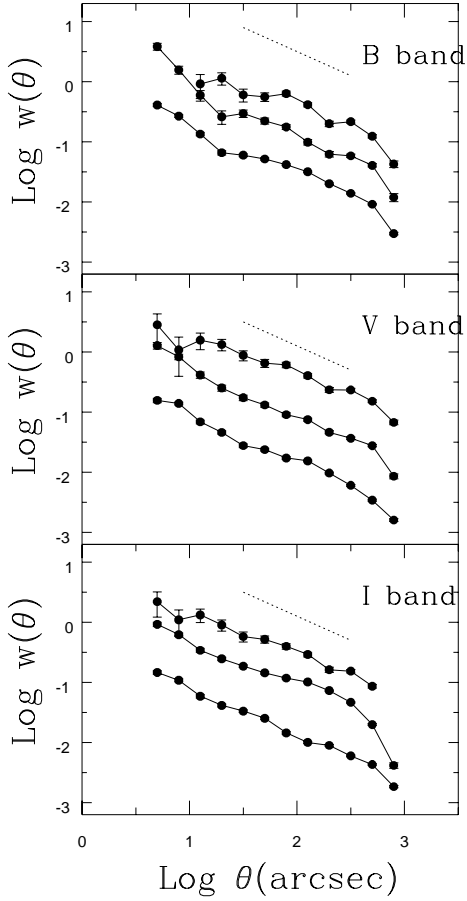}}
\caption[patchAevencor]{The angular two-point correlation function calculated for patch B, in
the B, V, and I bands as indicated. In each panel the three curves
correspond (from top to bottom) to the limiting magnitudes 21, 22, 24
(in B); 20, 22, 24 (in V); 19, 21, 23 (in I). The dotted line
represents a power law with a slope of $-0.8$.  The error bars are
1$\sigma$ errors obtained from 10 bootstrap realizations.  }
\label{fig:w}
\end{figure}

Figure~\ref{fig:w} shows $w(\theta)$ obtained for each of the three
passbands $B, V,$ and $I$, using the estimator proposed by Landy \&
Szalay (1993). The calculation has been done over the area defined
above (see figure~\ref{fig:proj_dist}). The error bars are $1\sigma$
errors calculated from ten bootstrap realizations. The angular
correlation function is, in general, well described by a power law
$\theta^{-{\gamma}}$ for angular scales extending out to $\theta\sim
0.5$ degrees, with a value of $\gamma$ in the range 0.7-0.8.  The
absence of any strong feature at the scale of the individual survey
frame should be noted. Furthermore, no significant variations of the
slope are detected, except at the bright end in all the three
passbands. In this case $w(\theta)$ is somewhat flatter than at
fainter magnitudes.  A possible explanation is the presence of the
nearby cluster (ACO S84) at $z
\sim 0.1$ located near the center of the patch. To test this
hypothesis the correlation function was recomputed by discarding a
square region of about 0.2 degrees on the side centered at the nominal
position of the cluster. Using the pruned sample yields a steeper
correlation function for the patch, consistent with the expected slope
of 0.8. These results show the uniformity of the EIS catalogs, once
obviously bad frames, selected on the basis of seeing and limiting
isophote, are discarded.

\begin{figure}
\resizebox{9cm}{!}{\includegraphics{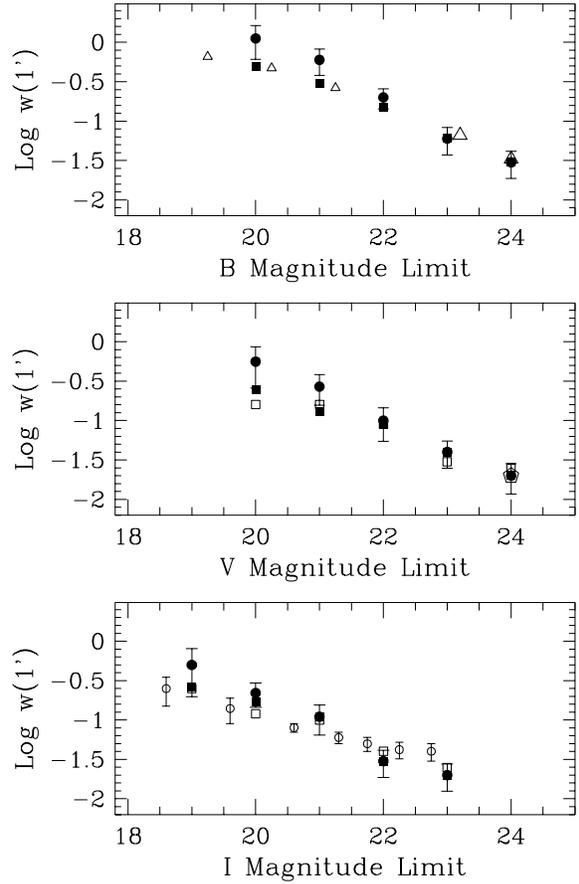}}
\caption[patchAevencor]{Amplitude of the correlation function measured at 1~arcmin for patch A
(open squares), and patch B (full circles) and after removing the
nearby S84 cluster (full squares). These results are compared to those
from Roche et al. (1993) (big open triangles) and Jones et al. (1987)
(small open triangles) in the B band, from Woods \& Fahlman (1997) in
the V band (open pentagon), and from Postman et al. (1998) in the I
band (open circles).  }
\label{fig:Aw}
\end{figure}

The angular correlation function can also be used to verify the
consistency of the photometric zero-points determined for patches A
and B. This can be done by studying the amplitude of the angular
correlation function as a function of limiting magnitude and comparing
the results obtained for the two patches.  Figure~\ref{fig:Aw} shows
the amplitude of the angular correlation function at a scale of
1~arcmin, $A_w$, as a function of the limiting magnitude in the
different passbands.  This amplitude is calculated from the best
linear-fits over the range $\sim$ 10-200 arcsec of $w(\theta)$ shown
in Figure~\ref{fig:w}. For $V$ and $I$ one finds good agreement
between the results for the two patches especially at the faint
end. The differences seen in the bright end are fully accounted for by
the presence of the nearby cluster described above.  The plot shows
the amplitude of the correlation with and without the cluster.  As can
be seen once the cluster is removed, the amplitude at the bright end
decreases and shows a good agreement with the estimate from patch A.
Similarly, one finds good agreement with the results obtained by other
authors such as: in the B band, Roche et al. (1993) ($23\le B
\le 24$) and Jones et al. (1987) ($19\le B \le 21$); in the V band,
Woods \& Fahlman (1997) (V=24); and in the I band, Postman et
al. (1998) ($19\le I \le 23$).  The amplitude of the angular
correlation function as determined from the EIS galaxy catalogs are
consistent with those obtained by various authors over the entire
range of magnitude. Note that at for $I \gsim 22$ the EIS points lie
slightly below those recently computed by Postman \etal (1998).

In summary, the above results are further evidence that the EIS galaxy
catalogs are uniform within a patch, that the zero-points for the
different patches are consistent and are in good agreement with external
data.

\section{Summary}

In this paper multicolor data for patch B covering about 1.7 square
degrees near the SGP have been presented. The quality of single band
and color catalogs have been assessed by comparisons of basic
statistics such as number counts, color distribution, and angular
correlation function, with model predictions and results from other
authors. The results indicate that the EIS object catalogs are
suitable for the science goals of the survey leading to consistent
results when compared to other data. Furthermore, one finds good
agreement between the results derived from the catalogs extracted from
patches A and B.

The work being carried out is essential in the preparation of the
final release of the EIS data. While the production of single-frame
catalogs is straightforward, the preparation of a catalog covering the
whole patch requires some experimentation in order to fine-tune the
parameters and  verify its uniformity, completeness, and reliability.
Multicolor data adds to the complexity of the task of catalog
production and further work in the preparation of well understood
color catalogs is required.  The experience gained so far points out
the need for a sustained effort in the development of techniques and
tools which will allow for: 1) the production of more customized
catalogs for different science goals, essential for public surveys; 2)
the exploration of the multi-dimensional space offered by multicolor
data; 3) the cross-correlation of the detected objects with the
increasing number of databases available in different wavelengths.
Searches in this multi-dimensional space offer unique science
opportunities and the implementation of suitable tools for its
exploration represent a major challenge for the efficient use of
imaging surveys carried out with the specific purpose of producing
targets to feed 8-m class telescopes.

The full range of products for patch A and B in the form of
astrometric and photometric calibrated pixel maps, object catalogs,
candidate target lists and on-line co-added section images can be
found at ``http://\-www.eso.org/eis/eis\_release.html''. New products will
be added incrementally as they become avaialble. Further information
on the project are available on the World Wide Web at
``http://\-www.eso.org/eis''.

\begin{acknowledgements}

We thank all the people directly or indirectly involved in the ESO
Imaging Survey effort. In particular, all the members of the EIS
Working Group for the innumerable suggestions and constructive
criticisms, the ESO Archive Group and ST-ECF for their support. We
also thank C. Lidman for providing his images, zero-points and
catalogs. We are also grateful to the DENIS consortium for making
available some of their survey data. The DENIS project development was
made possible thanks to the contributions of a number of researchers,
engineers and technicians in various institutes. The DENIS project is
supported by the SCIENCE and Human Capital and Mobility plans of the
European Commission under the grants CT920791 and CT940627, by the
French Institut National des Sciences de l'Univers, the Education
Ministry and the Centre National de la Recherche Scientifique, in
Germany by the State of Baden-Wurttemberg, in Spain by the DGICYT, in
Italy by the Consiglio Nazionale delle Richerche, by the Austrian
Fonds zur F\"orderung der wissenschaftlichen Forschung und
Bundesministerium f\"ur Wissenschaft und Forschung, in Brazil by the
Fundation for the development of Scientific Research of the State of
S\~ao Paulo (FAPESP), and by the Hungarian OTKA grants F-4239 and
F-013990 and the ESO C \& EE grant A-04-046.  Our special thanks to
A. Renzini, VLT Programme Scientist, for his scientific input, support
and dedication in making this project a success. Finally, we would
like to thank ESO's Director General Riccardo Giacconi for making this
effort possible.

\end{acknowledgements}

\end{document}